\documentclass[conference]{IEEEtran}
\IEEEoverridecommandlockouts
\usepackage{cite}
\usepackage{amsmath,amssymb,amsfonts}
\usepackage{algorithmic}
\usepackage{subcaption}
\usepackage{graphicx}
\usepackage{textcomp}
\usepackage{xcolor}
\def\BibTeX{{\rm B\kern-.05em{\sc i\kern-.025em b}\kern-.08em
    T\kern-.1667em\lower.7ex\hbox{E}\kern-.125emX}}

\usepackage{accents}
\newcommand{\ubar}[1]{\underaccent{\bar}{#1}}

\graphicspath{{./Images/}}

\begin{document}


\title{House Thermal Model Estimation: Robustness Across Seasons and Setpoints}

\author{
     Kunal Shankar\IEEEauthorrefmark{1}, Ninad Gaikwad\IEEEauthorrefmark{1}, and Anamika Dubey\IEEEauthorrefmark{1}\\
    
    \IEEEauthorblockA{\IEEEauthorrefmark{1}School of Electrical Engineering \& Computer Science, Washington State University, Pullman, USA} 
     
    \{ kunal.shankar, ninad.gaikwad, and anamika.dubey\}@wsu.edu
}

\maketitle


\begin{abstract}
Achieving the flexibility from house  heating, cooling,
and ventilation systems (HVAC) has the potential to enable large-scale demand response by aggregating HVAC load adjustments across many homes. This demand response strategy helps distribution grid to flexibly ramp-up or ramp-down local load demand so that it can optimally match the bulk power system generation profile. However, achieving this capability requires house thermal models that are both computationally efficient and robust to operating conditions. In this work, parameters of the Resistance-Capacitance (RC) network thermal model for houses are estimated using three optimization algorithms - Nonlinear Least Squares (NLS), Batch Estimation (BE), and Maximum Likelihood Estimation (MLE). The resulting models are evaluated through a Forward-Simulation across four different seasons and three setpoints. The results illustrate a principled way of selecting reduced order models and estimation methods with respect to the robustness offered to seasonal and setpoint variations in training-testing datasets.
 
\end{abstract}




\section{Introduction}\label{sec:Introduction}

Buildings contribute to approximately 37\% of total U.S. energy consumption, with residential houses alone accounting for 19.7\% according to the 2023 U.S. Energy Information Administration (EIA) statistics~\cite{EIAEnergyConsumptionWebpage}. Out of this share, a substantial portion of about $45\%$ is utilized by house HVAC systems. This positions houses as a promising asset for grid-edge applications, where they can provide ancillary services through demand response. Moreover, with the growing integration of renewable energy sources like solar and wind, harnessing the thermal inertia of these houses through the control of their HVAC systems present an effective approach to optimize grid operations. Realizing these objectives requires the development of grid-edge applications featuring coordinated control design of house HVAC systems at scale, making comprehensive simulation and analysis of these resources\cite{hu2017investigation}.

Coordinated control of house HVACs requires accurate, computationally efficient, and interpretable models of house thermal dynamics and HVAC power consumption. Achieving this depends critically on the choice of modeling approach. White-box models, such as EnergyPlus, provide detailed, physics-based representations that capture the underlying thermal processes, but are often computationally intensive, making scalability a challenge~\cite{foucquier2013state}. On the other hand, black-box models rely solely on data-driven methods and, while computationally lighter, typically lack the physical interpretability needed for model-based control design. Grey-box models bridge this gap by combining physics-based structure with data-driven estimation~\cite{li2021grey}, enabling computationally efficient yet physically meaningful modeling that supports scalable and effective grid-edge simulation and control design.
RC models are one of the most common approaches under grey-box modeling which makes use of equivalent electric circuits representing the fundamental physics principle of heat flowing from a higher temperature to lower temperature. This approach enables us to capture the thermal dynamics of houses \cite{wang2019predicting}. Parameter estimation for these models is challenging, as it involves solving non-linear optimization problems that can be formulated in several ways. Methods such as NLS~\cite{cui2019hybrid} and BE~\cite{kuhl2011real} have been used for parameter estimation in houses. While MLE has not, to our knowledge, been directly implemented for house thermal dynamics, its formulation has been addressed in related literature~\cite{simpson2023efficient}. Despite the prevalence of parameter estimation methods such as NLS, BE, and MLE in house thermal modeling, the existing literature seldom provides a comprehensive comparison of these techniques within a unified framework. Furthermore, few studies evaluate the ability of these models to accurately capture both temperature and HVAC power trajectories which gives us the amount of electricity the HVAC system is expected to utilize or examine the generalization of model performance across varying conditions.

In this work, we address these research gaps through the following contributions: (1) We present a unified and systematic comparative analysis of three commonly used parameter estimation methods (NLS, BE, and MLE) for RC-network thermal models of houses. (2) We evaluate both first-order (SM-1) and second-order (SM-2) RC models, enabling direct comparison across different levels of model complexity. (3) The predictive performance of each method and model is assessed based on the alignment of the thermal model and HVAC power model with observed data. (4) Finally, we provide a comprehensive assessment of robustness by testing all estimation methods and model structures across all four seasons and multiple HVAC setpoints. Collectively, these contributions offer practical guidance for selecting appropriate estimation techniques and model complexities to support robust, scalable house thermal modeling for grid-edge demand response applications.

The remainder of this paper is organized as follows. Section~II presents the modeling framework for grid-edge houses, focusing on the thermal dynamics of houses. Section~III introduces the thermal RC-network model, including the full four-state model used for data generation and the reduced-order models considered for parameter estimation. Section~IV details the model estimation methods, covering Nonlinear Least Squares, Batch Estimation, and Maximum Likelihood Estimation approaches. Section~V describes the case study setup along with the results and discussion, followed by the conclusions in Section~VI.


\section{House Model with HVAC Flexibility}\label{sec:Modeling}
We can have a generalized house model catering to applications of grid-edge which can be described as follows;
\begin{align}
    \ubar x(k+1) &=  f_{Build}(\ubar x(k), \ubar u(k), \ubar w(k) ; \ubar \theta), \label{eq:GM1} \\
    Q_{HVAC}(k) &=  f_{Q_{HVAC}}(\ubar x(k), \ubar u(k)), \label{eq:GM2} \\ 
     P_{HVAC}(k) &= f_{P_{HVAC}}(|Q_{HVAC}(k)|, \mathrm{COP}), \label{eq:GM3} \\
    P(k) &=   P_{HVAC}(k) +   P_{Other}(k), \label{eq:GM4} \\
    Q(k) &=  P(k)tan(cos^{-1}(pf)). \label{eq:GM5}
\end{align}

Equations~(\ref{eq:GM1})–(\ref{eq:GM5}) define the general house model. Equation~(\ref{eq:GM1}) describes the evolution of thermal states $\ubar{x}(k+1)$ as a function of their previous values, $\ubar{x}(k)$, control input, $\ubar{u}(k)$, disturbances, $\ubar{w}(k)$, and parameters. Equation~(\ref{eq:GM2}) gives the HVAC heat transfer rate, $Q_{HVAC}(k)$, which can be positive (heating) or negative (cooling) depending on the control action. Equation~(\ref{eq:GM3}) computes HVAC electrical power consumption,  $P_{HVAC}(k)$, from the heat transfer rate and coefficient of performance (COP). Total active power usage $P(k)$ in (\ref{eq:GM4}) is the sum of HVAC and other appliance loads. Finally, (\ref{eq:GM5}) computes the reactive power $Q(k)$ using $P(k)$ and the house’s power factor, $pf$s.

\section{Modeling of House Thermal Dynamics}\label{sec:Modeling}
\subsection{General Thermal RC-Network Model}\label{subsec:RCModeling}
The thermal dynamics of a house can be primarily represented using thermal RC network models which is based on basic physical property of heat flowing from higher-temperature areas to lower-temperature areas. These models resemble electrical circuits: nodes correspond to temperatures (analogous to voltage), resistors model thermal resistance to heat flow between nodes (paralleling electrical resistance to current), and capacitors represent thermal capacitance, reflecting the inertia in temperature change \cite{wang2019development}. A general model for RC network can shown as:
\begin{align}
    &C_z \frac{dT_z}{dt} = \sum_{i=1}^{N} \frac{T_{w_{i}} - T_{z}}{R_{zw_{i}}} + \frac{T_{am} - T_{z}}{R_{za}} + A_z Q_{HVAC} \label{eq:GB1} \\ 
    &\quad \quad \quad + B_z Q_{Int} + D_z Q_{Solar}, \nonumber \\
    & C_{w_{i}} \frac{dT_{w_{i}}}{dt} = \sum_{\substack{j=1 \\ j \neq i}}^{N} \frac{T_{w_{j}} - T_{w_{i}}}{R_{w_{ij}}} + \frac{T_z - T_{w_{i}}}{R_{zw_{i}}} + \frac{T_{am} - T_{w_{i}}}{R_{wa_{i}}} \label{eq:GB2} \\
    &\quad \quad \quad + B_{w_{i}} Q_{Int} + D_{w_{i}} Q_{Solar}. \nonumber 
\end{align}

$T_{z}$ is the average temperature of a house, $T_{w_{i}}$ is $i^{\text{th}}$ unmeasured state out of $N$ (typically associated with fictitious internal surfaces used to represent heat transfer dynamics) and $T_{am}$ is the outside ambient temperature.  $Q_{Solar}$, $Q_{Int}$ and $Q_{HVAC}$ are the heat gains from solar irradiance, internal heat appliances/sources and heat accepted or rejected due to cooling or heating by HVAC respectively. $A_z$, $B_z$-$B_{w_{i}}$, $D_z$-$D_{w_{i}}$ are the factors associated with the various heat gains which basically shows the effect of these heat gains on the states $T_{z}$ and $T_{w}$.

We can write the eq.s (\ref{eq:GB1}) and (\ref{eq:GB2}) as a general linear state-space model given as;
\begin{align}
    \dot{\ubar{x}} &=  A( \ubar \theta) \ubar{x} + B(\ubar \theta) \ubar{u} + D(\ubar \theta) \ubar{w} \label{eq:LSS1} \\
    y &= C \ubar{x} \label{eq:LSS2}.
\end{align}
The state vector is defined as $\ubar{x} = [T_{z}, T_{w_{1}}, \dots, T_{w_{N}}]^{T}$, with control input $\ubar{u} = Q_{HVAC}$, and disturbance vector $\ubar{w} = [T_{am}, Q_{Int}, Q_{Solar}]^{T}$. The system output is $y = T_{z}$, corresponding to the only measurable state. The matrices $A(\ubar{\theta})$, $B(\ubar{\theta})$, and $D(\ubar{\theta})$ denote the system, input, and disturbance matrices, respectively, parameterized by $\ubar{\theta}$.

The parameter vector is given by:
\begin{equation*}
\theta = \begin{bmatrix}
\begin{aligned}
&R_{za}, R_{zw_1}, \dots, R_{zw_N}, R_{w_{1,2}}, \dots, R_{w_{N-1,N}}, \\
&R_{wa_1}, \dots, R_{wa_N}, C_z, C_{w_1}, \dots, C_{w_N}, \\
&A_z, B_z, B_{w_1}, \dots, B_{w_N}, D_z, D_{w_1}, \dots, D_{w_N}
\end{aligned}
\end{bmatrix}^{T}
\end{equation*}

\noindent which encapsulates the thermal resistances, capacitance, and the coefficients associated with the various heat gains. These parameters are to be estimated from measured data. The output matrix is defined as $C = [1, 0, \dots, 0]$, reflecting that only the indoor air temperature $T_{z}$ is observable. It is important to note that the complexity and expressive power of the model grow with $N$. Under a suitable time discretization scheme, ~\eqref{eq:GB1} leads to the discrete-time representation of house thermal model, $f_{Build}(\ubar{x}(k), \ubar{u}(k), \ubar{w}(k); \ubar{\theta})$, used in the generalized house model in Section II. This model can further be tailored to different building archetypes by adjusting the number of thermal states and the corresponding network of resistances and capacitances,

\subsection{Four-State RC Model for Data Generation (SM4)}\label{subsec:4StateModel}

To simulate realistic house thermal behavior, we adopt a physics-inspired four-state RC network model. This model captures the dominant heat transfer pathways between the indoor air, attic space, and internal mass, each represented as a temperature state \cite{cui2019hybrid}. The governing differential equations are given by:
\begin{align}
    C_{in} \frac{dT_{z}}{dt} &= \frac{T_{wall} - T_{z}}{R_{w}/2} 
                   + \frac{T_{attic} - T_{z}}{R_{attic}} 
                   + \frac{T_{im} - T_{z}}{R_{im}} \nonumber \\
                   &\quad + \frac{T_{am} - T_{z}}{R_{win}} 
                   + A_{in}Q_{IHL} +B_{in}Q_{AC} \label{eq:4state_ave} \\
                   C_w \frac{dT_{wall}}{dt} &= \frac{1}{R_{w}/2}(({T_{sol,w} - T_{wall}}) - ({T_{wall} - T_{z}})) \label{eq:4state_wall} \\
    C_{attic} \frac{dT_{attic}}{dt} &= \frac{T_{sol,r} - T_{attic}}{R_{roof}} - \frac{T_{attic} - T_{z}}{R_{attic}} \label{eq:4state_attic} \\
    C_{im} \frac{dT_{im}}{dt} &= \frac{T_{im} - T_{z}}{R_{im}} + D_{im}Q_{solar} \label{eq:4state_im}
\end{align}

Here, $T_{z}$ denotes the measurable indoor air temperature, while $T_{wall}$, $T_{attic}$, and $T_{im}$ represent the unmeasurable temperatures of the wall mass, attic space, and internal mass, respectively. $T_{sol,w}$ and $T_{sol,r}$ are solar-equivalent surface temperatures affecting the wall and roof, and $T_{am}$ is the ambient outdoor temperature. $Q_{solar}$, $Q_{IHL}$, and $Q_{AC}$ represent the heat gains from solar irradiance, internal loads (e.g., appliances and occupants), and HVAC heating or cooling, respectively along with coefficients $A_{in}$, $B_{in}$ and $D_{im}$ capturing their relative influence on the thermal states. $R_{w}$, $R_{attic}$, $R_{im}$, $R_{win}$, and $R_{roof}$ represent the thermal resistances between the respective nodes, while $C_{w}$, $C_{in}$, $C_{attic}$, and $C_{im}$ denote the thermal capacitance of the wall, indoor air, attic, and internal mass, respectively.

In this paper, this four-state model is used to generate the synthetic data. The structure and parameters given in \cite{cui2019hybrid}, this allows us to treat them as \emph{ground truth} for evaluating the accuracy of reduced-order models hence  allowing for direct comparison of estimated parameters against true values.

\subsection{Reduced-Order Models}\label{subsec:ReducedOrderModels}

In order to enable data-driven modeling and reduce computational complexity, we consider two simplified RC models: a second-order model (SM2) and a first-order model (SM1). These models are trained using data generated from the 4-state physical model (SM4) discussed previously.

\subsubsection{Second-Order Model (SM2)}
\begin{align}
    C_{in_{sm2}} \frac{dT_{z}}{dt} &= 
    \frac{T_{wall_{sm2}} - T_{z}}{R_{w_{sm2}}/2}
    + \frac{T_{am_{sm2}} - T_{z}}{R_{win_{sm2}}} \nonumber \\[4pt]
    &\quad + A_{ih_{sm2}} Q_{ih}
    + B_{ac_{sm2}} Q_{ac} \label{eq:sm2_ave} \\[8pt]
    C_{w_{sm2}} \frac{dT_{wall_{sm2}}}{dt} &=
    \frac{T_{z} - T_{wall_{sm2}}}{R_{w_{sm2}}/2} - \frac{T_{sol,w_{sm2}} - T_{wall_{sm2}}}{R_{w_{sm2}}/2}\nonumber \\[4pt]
    &\quad 
    + D_{solar_{sm2}} Q_{solar} \label{eq:sm2_wall}
\end{align}

The SM2 model describes the evolution of two thermal states: the indoor air temperature $T_{z}$ and the wall temperature $T_{wall_{sm2}}$. The dynamics are governed by heat exchange with the ambient environment ($T_{am_{sm2}}$) and the solar-facing wall surface ($T_{sol,w_{sm2}}$) through resistances $R_{win_{sm2}}$ and $R_{w_{sm2}}$. Internal heat gains, HVAC operation, and solar radiation are represented by $Q_{ih}$, $Q_{ac}$, and $Q_{solar}$, scaled by the gain coefficients $A_{ih_{sm2}}$, $B_{ac_{sm2}}$, and $D_{solar_{sm2}}$, respectively. The system’s thermal storage is captured by the capacitances $C_{in_{sm2}}$ and $C_{w_{sm2}}$.

\subsubsection{First-Order Model (SM1)}
\begin{align}
    C_{in_{sm1}} \frac{dT_{z}}{dt} &=
    \frac{T_{am_{sm1}} - T_{z}}{R_{win_{sm1}}}
    + A_{ih_{sm1}} Q_{ih}
    + B_{ac_{sm1}} Q_{ac} \nonumber \\[4pt]
    &\quad + D_{solar_{sm1}} Q_{solar} \label{eq:sm1_ave}
\end{align}

The SM1 model captures single-zone thermal dynamics by modeling the indoor air temperature $T_{z}$ as the sole state. Heat transfer occurs between the zone and the ambient air $T_{am_{sm1}}$, governed by the resistance $R_{win_{sm1}}$. In addition, the model incorporates thermal inputs from internal sources ($Q_{ih}$), HVAC equipment ($Q_{ac}$), and solar radiation ($Q_{solar}$), each scaled by the respective gains $A_{ih_{sm1}}$, $B_{ac_{sm1}}$, and $D_{solar_{sm1}}$. The zone’s capacity to store heat is represented by $C_{in_{sm1}}$.

The SM4, SM2, and SM1 models introduced above can all be expressed in the general state-space form given in (\ref{eq:LSS1})–(\ref{eq:LSS2}). The state vector $\ubar{x}$ consists of four, two, and one thermal states for SM4, SM2, and SM1, respectively: $\ubar{x}_{\text{SM4}} = [T_{z}, T_{wall}, T_{attic}, T_{im}]^{T}$, $\ubar{x}_{\text{SM2}} = [T_{z}, T_{wall}]^{T}$, and $\ubar{x}_{\text{SM1}} = [T_{z}]$.The input vector is $\ubar{u} = [Q_\mathrm{ac}]$ for all models.The disturbance vector is defined as $\ubar{w}_{\text{SM4}} = [Q_{IHL}, Q_{solar}, T_{sol,w}, T_{sol,r}, T_{am}]^{T}$, $\ubar{w}_{\text{SM2}} = [Q_{IHL}, Q_{solar}, T_{sol,w}, T_{am}]^{T}$, and $\ubar{w}_{\text{SM1}} = [Q_{IHL}, Q_{solar}, T_{am}]^{T}$.The system output is $y = T_{z}$, which corresponds to the only measurable state in all models. The matrices $A(\ubar{\theta})$, $B(\ubar{\theta})$, and $D(\ubar{\theta})$ represent the system, input, and disturbance dynamics, respectively, and are parameterized by the vector $\ubar{\theta}$.
.The parameter vector $\ubar{\theta}$ encapsulates thermal resistances, capacitances, and gain coefficients, with dimensionality varying by model: 12 parameters for SM4, 7 for SM2, and 5 for SM1. Specifically, $\ubar{\theta}_{\text{SM4}} = [R_{w}, R_{attic}, R_{im}, R_{win}, R_{roof}, C_{in}, C_{w}, C_{attic}, C_{im}, A_{ih}, B_{ac}, \\ D_{solar}]^{T}$. For SM2, $\ubar{\theta}_{\text{SM2}} = [R_{w}, R_{win}, C_{in}, C_{w}, A_{ih}, B_{ac}, D_{solar}]^{T}$. For SM1, $\ubar{\theta}_{\text{SM1}} = [R_{win}, C_{in}, A_{ih}, B_{ac}, D_{solar}]^{T}$. This structure allows each model to be interpreted under a unified linear parameterized framework, with increasing model complexity reflected by the number of states and parameters.


\section{Model Estimation Methods}\label{sec:Methods}
In order to estimate the ($\ubar{\theta}$): parameters of the model, we make use of the $T$ samples which we collect in the input output data tuples as $\mathbb{D} = \{ (\ubar{u}(1), \ubar{w}(1), y(1)), \dots, (\ubar{u}(T), \ubar{w}(T), y(T)) \}$ where the inputs to our system is  $\ubar{u}$ and $\ubar{w}$. The measured output is represented by $y$.


\subsection{Estimation Methods for Thermal RC-Network Model}\label{subsec:RCEstimationMethods}
Association of parameter vector with ~(\ref{eq:GB1}) and ~(\ref{eq:GB2}) makes it non-linear. Using the standard method of Euler discretization for the continuous-time dynamics combined along with three commonly used nonlinear programming formulations, we can estimate the  $\ubar{\theta}$ . The three methods are as follows:

\subsubsection{Nonlinear Least Squares}\label{subsubsec:NLSMethod}

The Nonlinear Least Squares (NLS) approach represents the most basic parameter estimation method, as formulated in eqs.~(\ref{eq:LS_objective})--(\ref{eq:LS_output}). The objective function minimizes the sum of squared errors between the measured output and the model-predicted output, subject to the thermal RC-network dynamics.

\begin{align}
    & \min_{\ubar x, \ubar \theta} \quad \sum_{k=1}^{T} \left( y(k) - \tilde{y}(k) \right)^2 \label{eq:LS_objective} \\
    &\text{subject to:}  \nonumber \\
    & \ubar{x}(k+1) = \ubar{x}(k) + t_s [ A( \ubar \theta) \ubar{x}(k) + B(\ubar \theta) \ubar{u}(k)  \label{eq:LS_dynamics} \\
    & \quad \quad \quad \quad + D(\ubar \theta) \ubar{w}(k) ], \nonumber \\
    & \tilde{y}(k) = C \ubar{x}(k) .  \label{eq:LS_output}
\end{align}

\subsubsection{Batch Estimation}\label{subsubsec:BEMethod}

The Batch Estimation (BE) formulation, shown in eqs.~(\ref{eq:BE_objective})--(\ref{eq:BE_residual}), accounts for both process noise—arising from mismatches between the model structure and the true system—and measurement noise due to sensor inaccuracies. For further details, refer to~\cite{robertson1996moving}.
\begin{align}
    & \min_{\ubar w_{n}, v_{n}, \ubar x, \ubar \theta} \quad (\ubar x^{e}_{0})^T P_0^{-1} \ubar x^{e}_{0} + \sum_{k=0}^{T} \left( v_{n}(k)^T R^{-1} v_{n}(k) \right) \label{eq:BE_objective} \\ 
    & \quad \quad \quad + \sum_{k=0}^{T-1} \left( \ubar w_{n}(k)^T Q^{-1} \ubar w_{n}(k) \right) \nonumber \\
    & \text{subject to:}  \nonumber \\    
    & \ubar x(k+1) = \ubar{x}(k) + t_s [A( \ubar \theta) \ubar{x}(k) + B(\ubar \theta) \ubar{u}(k) \label{eq:BE_state_dynamics} \\
    & \quad \quad \quad \quad + D(\ubar \theta) \ubar{w}(k) ] + \ubar w_{n}(k), \nonumber \\
    & v_{n}(k) = y(k) - C \ubar{x}(k) . \label{eq:BE_residual} 
\end{align}

Here, $\ubar{x}^{e}_{0} \triangleq \ubar{x}(0) - \ubar{x}_{0}$ denotes the initial state estimation error, where $\ubar{x}_{0}$ is the initial state estimate, and $P_0$ represents the corresponding error covariance matrix. The process noise is modeled as $\ubar{w}_{n} \sim \mathcal{N}(\ubar{0}, Q)$ with covariance matrix $Q$, and the measurement noise as $v_{n} \sim \mathcal{N}(0, R)$ with variance $R$. In practice, $Q$ and $R$ are treated as hyperparameters and are tuned based on the estimation performance.

\subsubsection{Maximum Likelihood Estimation}\label{subsubsec:MLEMethod}

The Maximum Likelihood Estimation (MLE) approach, like the Batch Estimation (BE) method, accounts for both process and measurement noise. However, it does so by incorporating a Kalman Filter-based formulation of the thermal RC-network dynamics within the constraints, and by minimizing the one-step-ahead prediction error ($e$) obtained during the filtering process. The complete MLE formulation is given in eqs.~(\ref{eq:MLE_objective})--(\ref{eq:MLE_S}). For further details, see~\cite{simpson2023efficient}.
\begin{align}
    & \min_{P, S, \ubar e, \ubar{\tilde{x}}, \ubar \theta} \quad 
   \sum_{k=1}^{T} \ubar e(k)^{T} S(k)^{-1} \ubar e(k) + \log \mid S(k) \mid \label{eq:MLE_objective}\\
    & \text{subject to:}  \nonumber \\
    & \ubar{\tilde{x}}(k+1) = (I + t_{s} I) A( \theta ) [ \ubar{ \tilde{x} }(k) + P(k) C^T S(k)^{-1}  ) \label{eq:MLE_state_update}\\
    & \quad \quad  \quad \quad  \quad e(k) ] + t_{s} B(\theta) \ubar{u(k)} +   D(\theta) \ubar{w(k)} ], \nonumber \\
    & C(k+1) = (I + t_{s} I) A(\theta) [ P(k) - P(k) C^T S(k)^{-1}  \label{eq:MLE_covariance_update} \\
    & \quad \quad  \quad \quad  \quad  C P(k)] (I + t_{s} I) A(\theta)^T + Q, \nonumber  \\
    & \ubar e(k) = y(k) - C \ubar{\tilde{x}}(k), \label{eq:MLE_residual} \\
    & S(k) = C P(k) C^T + R \label{eq:MLE_S}.
\end{align}
Where, $\ubar{ \tilde{x} }$, $P$ and $S$ are the predicted state, state error covariance matrix, and the one-step error variance respectively.

\section{Case Study}\label{sec:CaseStudy}
\begin{figure*}[ht]
    \centering

    \begin{subfigure}[t]{0.16\textwidth}
        \centering
        \includegraphics[width=\textwidth]{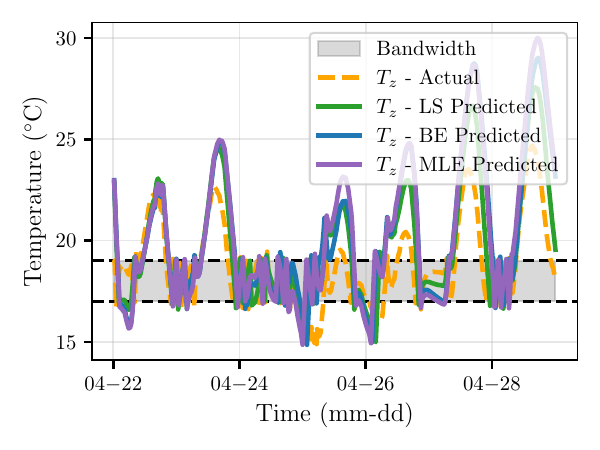}
        \caption{SM-1 Spring 18\textdegree{}C}
    \end{subfigure}
    \begin{subfigure}[t]{0.16\textwidth}
        \centering
        \includegraphics[width=\textwidth]{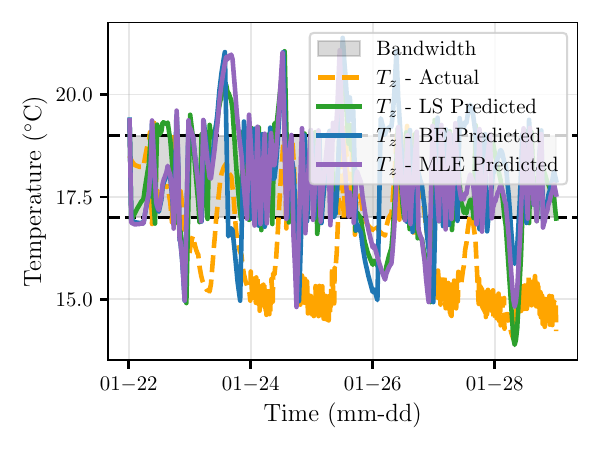}
        \caption{SM-1 Winter 18\textdegree{}C}
    \end{subfigure}
    \begin{subfigure}[t]{0.16\textwidth}
        \centering
        \includegraphics[width=\textwidth]{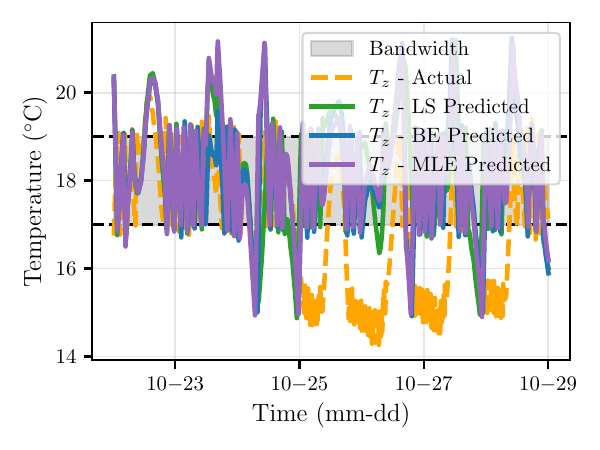}
        \caption{SM-1 Fall 18\textdegree{}C}
    \end{subfigure}
    \begin{subfigure}[t]{0.16\textwidth}
        \centering
        \includegraphics[width=\textwidth]{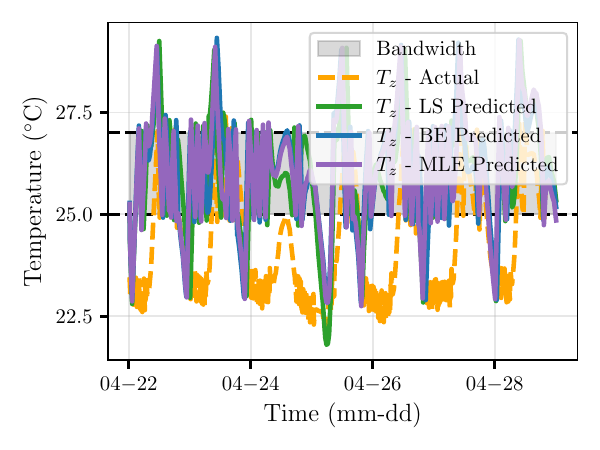}
        \caption{SM-1 Spring 26\textdegree{}C}
    \end{subfigure}
    \begin{subfigure}[t]{0.16\textwidth}
        \centering
        \includegraphics[width=\textwidth]{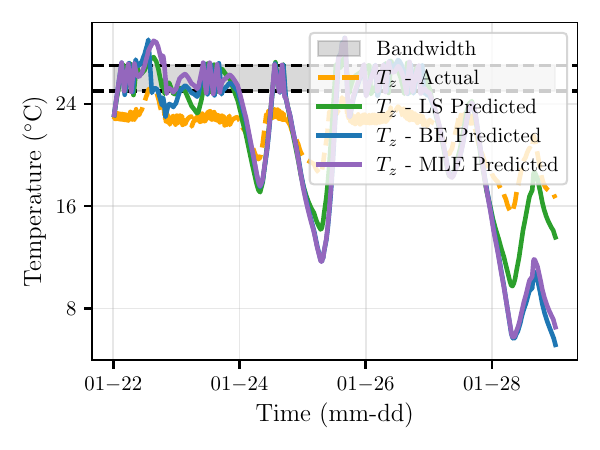}
        \caption{SM-1 Winter 26\textdegree{}C}
    \end{subfigure}
    \begin{subfigure}[t]{0.16\textwidth}
        \centering
        \includegraphics[width=\textwidth]{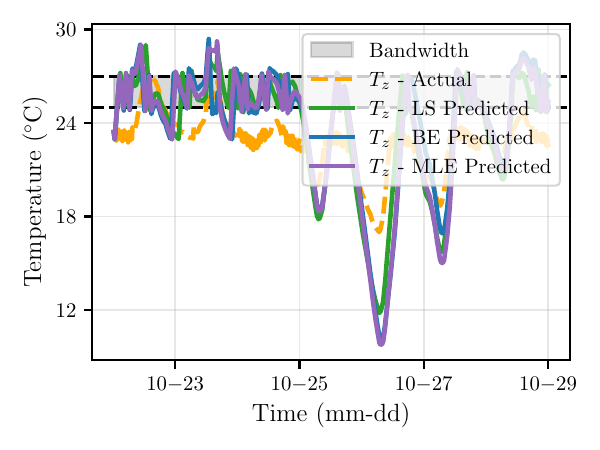}
        \caption{SM-1 Fall 26\textdegree{}C}
    \end{subfigure}

    \begin{subfigure}[t]{0.16\textwidth}
        \centering
        \includegraphics[width=\textwidth]{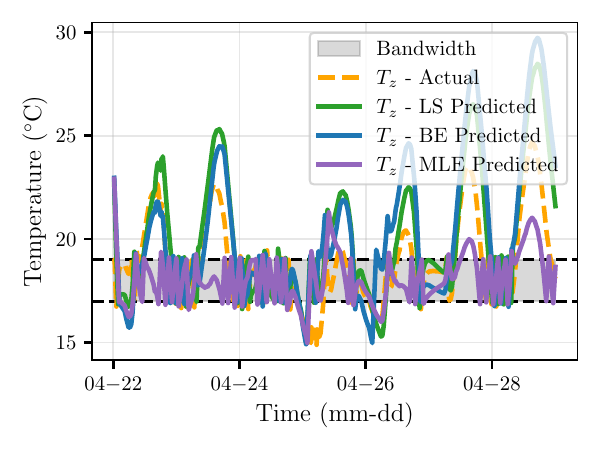}
        \caption{SM-2 Spring 18\textdegree{}C}
    \end{subfigure}
    \begin{subfigure}[t]{0.16\textwidth}
        \centering
        \includegraphics[width=\textwidth]{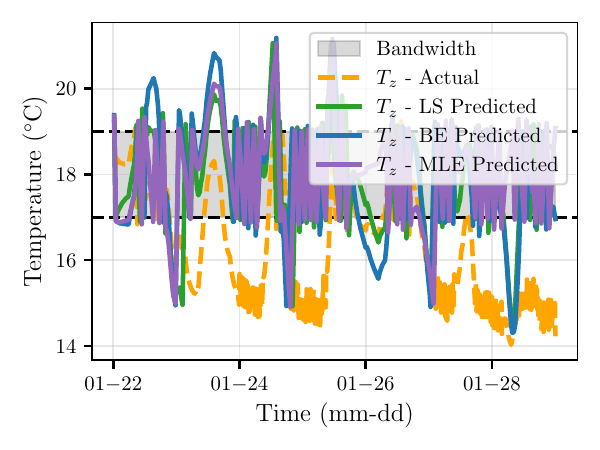}
        \caption{SM-2 Winter 18\textdegree{}C}
    \end{subfigure}
    \begin{subfigure}[t]{0.16\textwidth}
        \centering
        \includegraphics[width=\textwidth]{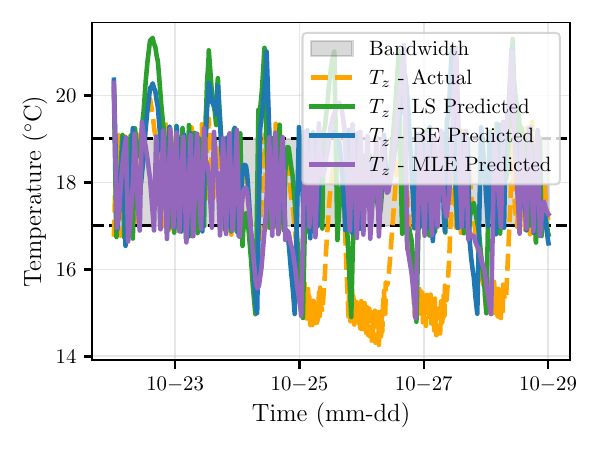}
        \caption{SM-2 Fall 18\textdegree{}C}
    \end{subfigure}
    \begin{subfigure}[t]{0.16\textwidth}
        \centering
        \includegraphics[width=\textwidth]{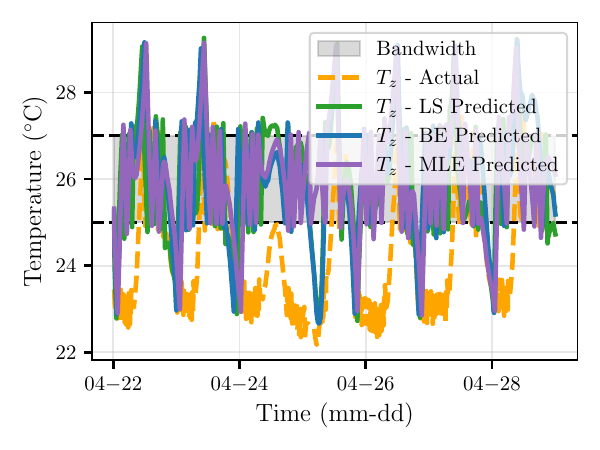}
        \caption{SM-2 Spring 26\textdegree{}C}
    \end{subfigure}
    \begin{subfigure}[t]{0.16\textwidth}
        \centering
        \includegraphics[width=\textwidth]{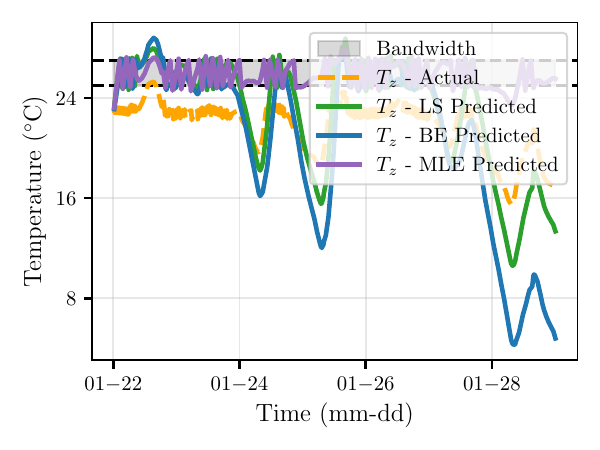}
        \caption{SM-2 Winter 26\textdegree{}C}
    \end{subfigure}
    \begin{subfigure}[t]{0.16\textwidth}
        \centering
        \includegraphics[width=\textwidth]{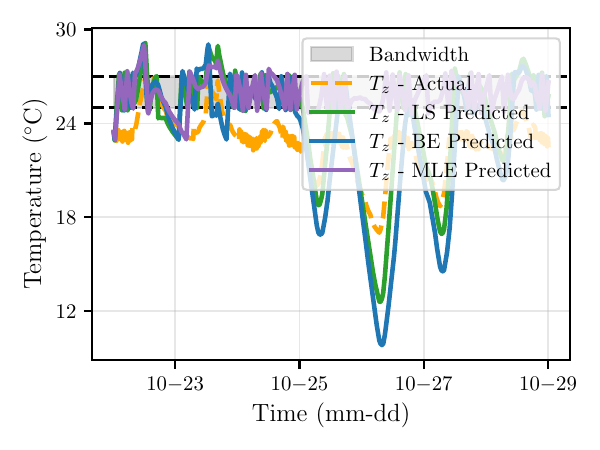}
        \caption{SM-2 Fall 26\textdegree{}C}
    \end{subfigure}

    \begin{subfigure}[t]{0.16\textwidth}
        \centering
        \includegraphics[width=\textwidth]{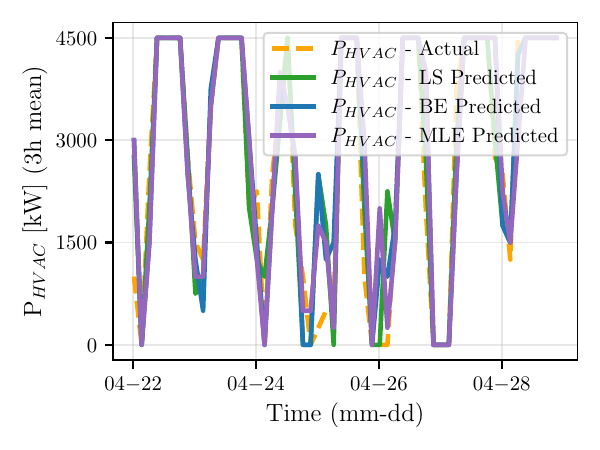}
        \caption{SM-1 Spring 18\textdegree{}C}
    \end{subfigure}
    \begin{subfigure}[t]{0.16\textwidth}
        \centering
        \includegraphics[width=\textwidth]{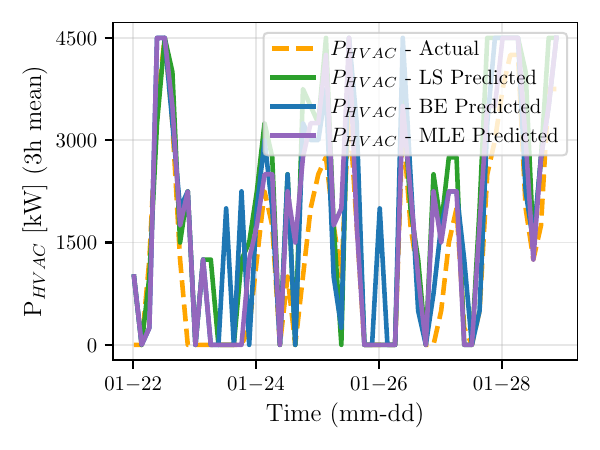}
        \caption{SM-1 Winter 18\textdegree{}C}
    \end{subfigure}
    \begin{subfigure}[t]{0.16\textwidth}
        \centering
        \includegraphics[width=\textwidth]{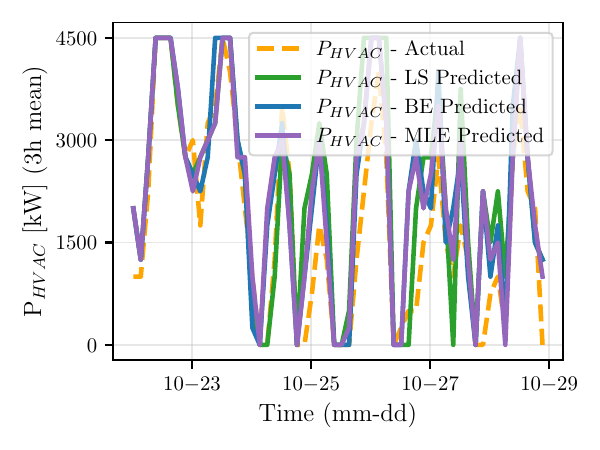}
        \caption{SM-1 Fall 18\textdegree{}C}
    \end{subfigure}
    \begin{subfigure}[t]{0.16\textwidth}
        \centering
        \includegraphics[width=\textwidth]{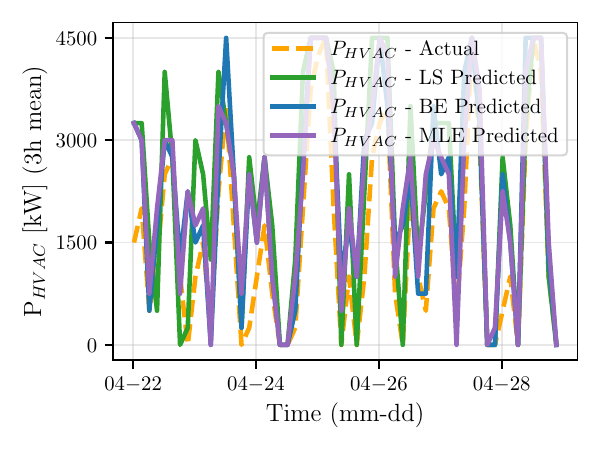}
        \caption{SM-1 Spring 26\textdegree{}C}
    \end{subfigure}
    \begin{subfigure}[t]{0.16\textwidth}
        \centering
        \includegraphics[width=\textwidth]{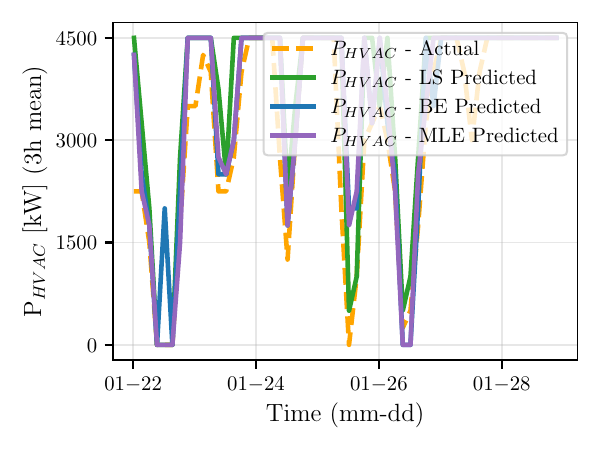}
        \caption{SM-1 Winter 26\textdegree{}C}
    \end{subfigure}
    \begin{subfigure}[t]{0.16\textwidth}
        \centering
        \includegraphics[width=\textwidth]{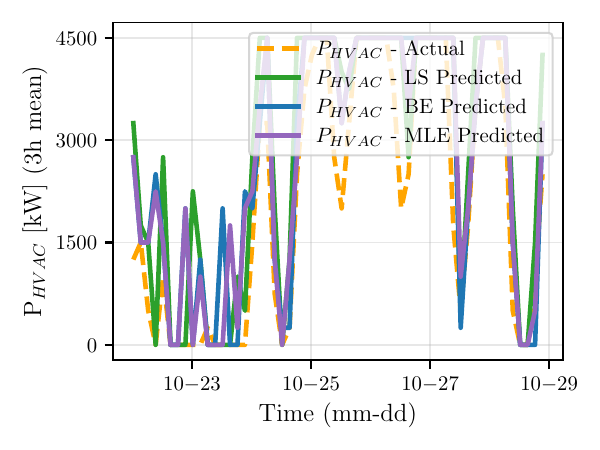}
        \caption{SM-1 Fall 26\textdegree{}C}
    \end{subfigure}

    \begin{subfigure}[t]{0.16\textwidth}
        \centering
        \includegraphics[width=\textwidth]{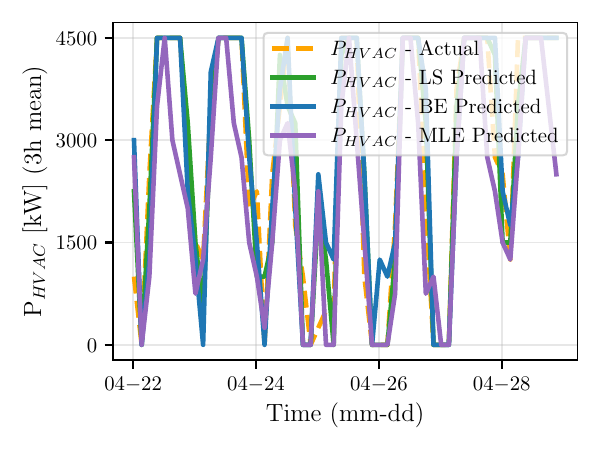}
        \caption{SM-2 Spring 18\textdegree{}C}
    \end{subfigure}
    \begin{subfigure}[t]{0.16\textwidth}
        \centering
        \includegraphics[width=\textwidth]{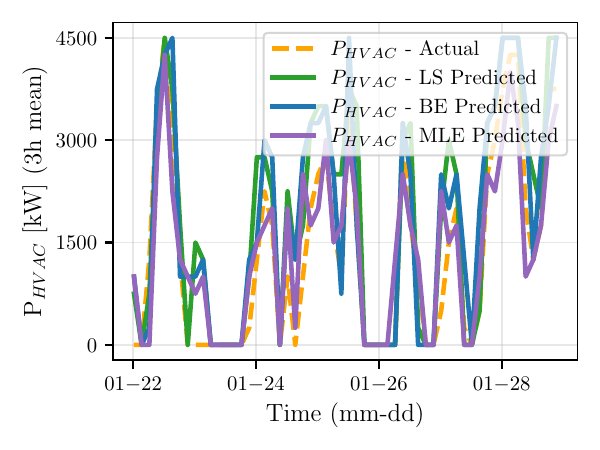}
        \caption{SM-2 Winter 18\textdegree{}C}
    \end{subfigure}
    \begin{subfigure}[t]{0.16\textwidth}
        \centering
        \includegraphics[width=\textwidth]{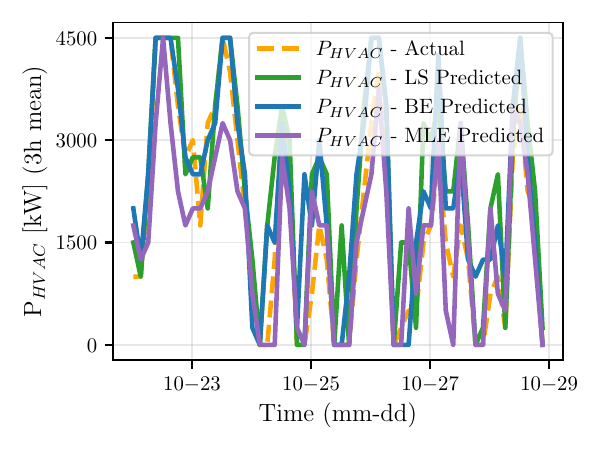}
        \caption{SM-2 Fall 18\textdegree{}C}
    \end{subfigure}
    \begin{subfigure}[t]{0.16\textwidth}
        \centering
        \includegraphics[width=\textwidth]{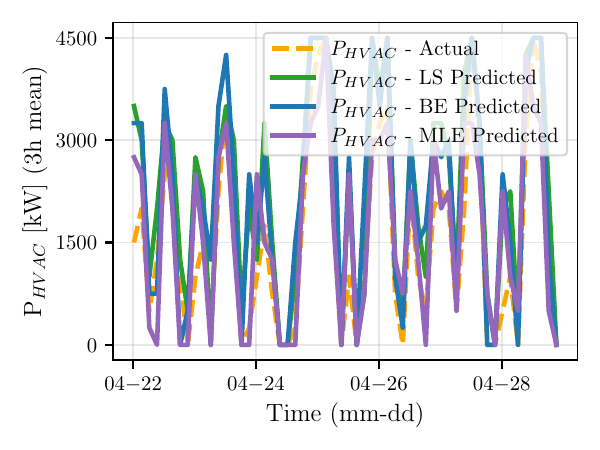}
        \caption{SM-2 Spring 26\textdegree{}C}
    \end{subfigure}
    \begin{subfigure}[t]{0.16\textwidth}
        \centering
        \includegraphics[width=\textwidth]{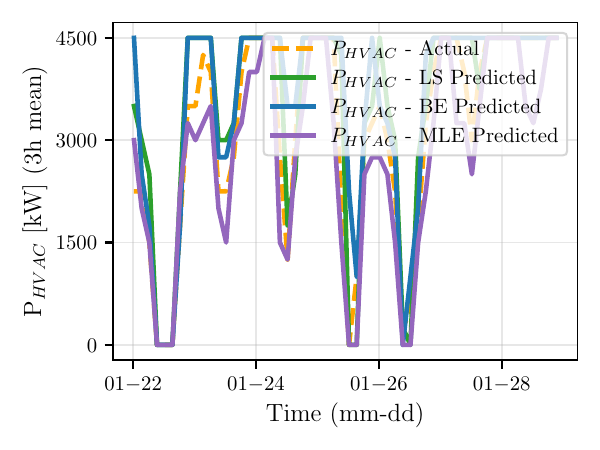}
        \caption{SM-2 Winter 26\textdegree{}C}
    \end{subfigure}
    \begin{subfigure}[t]{0.16\textwidth}
        \centering
        \includegraphics[width=\textwidth]{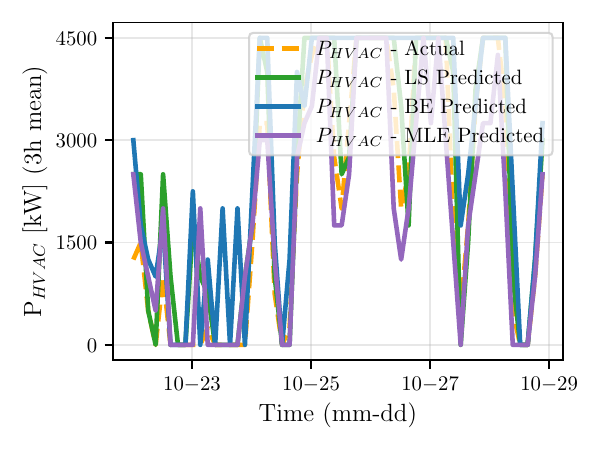}
        \caption{SM-2 Fall 26\textdegree{}C}
    \end{subfigure}

    \caption{Performance comparison of $T_{z}$ and $P_{HVAC}$ trajectories across different testing periods (spring, winter, fall) and setpoints (18\textdegree{}C, 26\textdegree{}C).}
    \vspace{-0.3cm}
    \label{fig:Temp_PHVAC_SM1_SM2}
\end{figure*}

\begin{figure}[htbp]
    \centering
    \begin{subfigure}[t]{0.98\linewidth}
        \centering
        \includegraphics[width=0.95\linewidth,height=0.28\textheight,keepaspectratio]{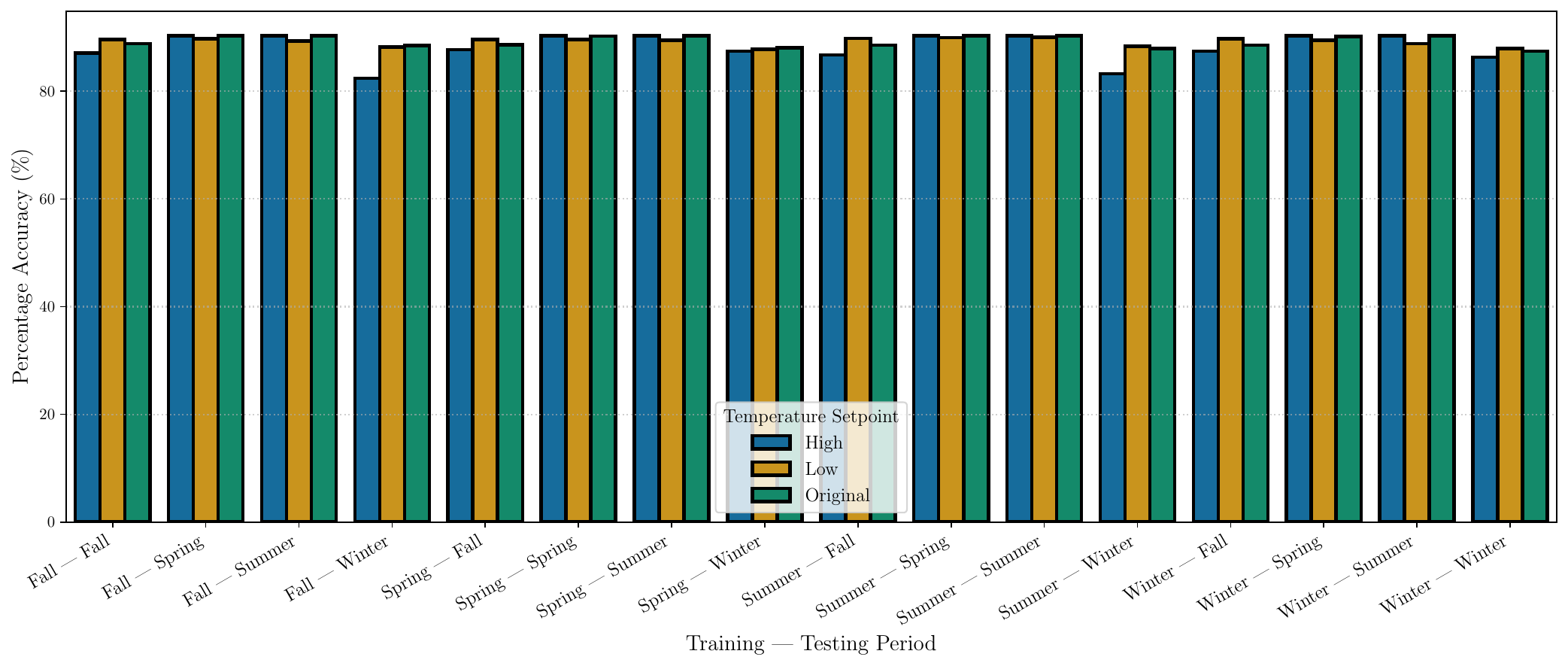}
        \caption{$T_{z}$: Train + Test Period + Setpoint}
        \label{fig:TEMP_Train_Test_Period}
    \end{subfigure}
    \vspace{0.7em}

    \begin{subfigure}[t]{0.98\linewidth}
        \centering
        \includegraphics[width=0.95\linewidth,height=0.28\textheight,keepaspectratio]{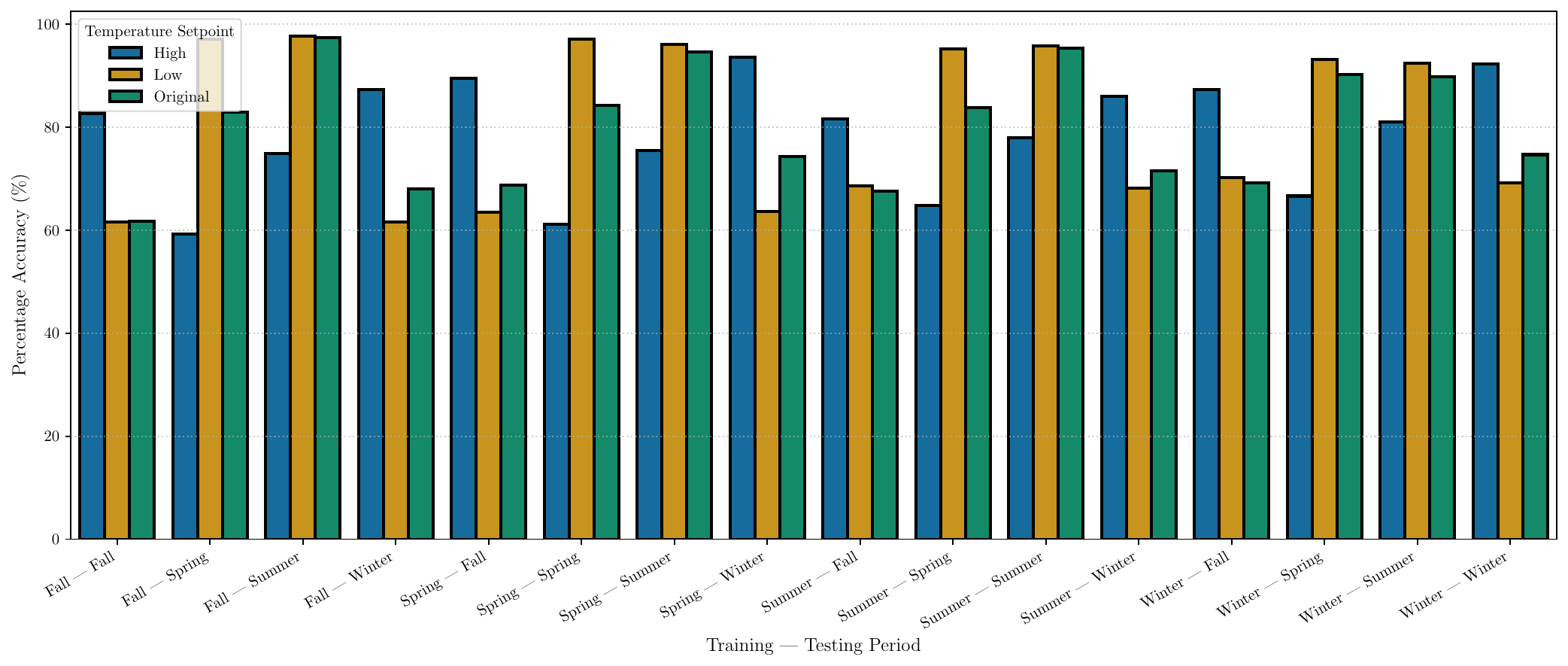}
        \caption{$P_{HVAC}$: Train + Test Period + Setpoint}
        \label{fig:PHVAC_Train_Test_Period}
    \end{subfigure}

    \caption{Comparison of average accuracy for $T_{z}$ and $P_{HVAC}$ trajectories across training and testing periods.}
    \label{fig:TEMP_PHVAC_Accuracy_2x1}
\end{figure}

\subsection{Setup}\label{sec:Setup}\textbf{}

\subsubsection{Data Generation}\label{subsubsec:Data Generation}

Data is generated at 10-minute intervals using a fourth-order ODE simulation of a single-family home’s thermal dynamics. The simulation uses real inputs: household electrical demand from the Pecan Street dataset (Austin, TX) and weather data from NSRDB (Gainesville, FL), both covering all of 2017 at 10-minute resolution. Simulations are performed for three HVAC setpoints: 18$^\circ$C, 22$^\circ$C, and 26$^\circ$C.
\subsubsection{Training Data}\label{subsubsec:ForwardSim}



A full year of data is generated using the fourth-order thermal model at a fixed HVAC setpoint of 22$^\circ$C for training. This dataset is divided into four seasonal quarters of Winter (January–March), Spring (April–June), Summer (July–September), and Fall (October–December). For each season, SM1 and SM2 models are trained using a 21-day data window with three optimization methods (LS, MLE, BE), resulting in $2 \times 3 \times 4 = 24$ unique parameter sets. Extending the training data beyond 21 days did not significantly improve predictions but increased computational costs and caused convergence issues. These sets are used for all subsequent forward simulations and testing.

\subsubsection{Testing Data}\label{subsubsec:TestingData}

To evaluate robustness, each parameter set is trained on a specific quarter at 22$^\circ$C is tested across all four seasonal quarters and three HVAC setpoints (18$^\circ$C, 22$^\circ$C, 26$^\circ$C), yielding $4 \times 3 = 12$ testing scenarios per set for both models and all optimization methods. In forward simulation, these parameters are fixed and used to simulate the house’s thermal response over a 30-day test window across seasons and setpoints. Zero$^{\mathrm{th}}$-Order-Hold integration is used for forward simulation where states are fed back at each step, and a local controller regulates the indoor environment for the different setpoints. By comparing predicted indoor temperature and $P_\mathrm{HVAC}$ to ground truth, we assess how well the model generalizes to unseen conditions. We use the performance metric of Percentage Accuracy = 100 - Mean Absolute Percentage Error (MAPE).

\subsubsection{Computation}\label{subsubsec:Computation}  
The estimation methods for RC-Network models are formulated as nonlinear programs in CasADi~\cite{CasADi} and solved using IPOPT~\cite{Ipopt}.

\subsection{Results and Discussion}
\label{sec:ResultsDiscussion}

\subsubsection{$T_{z}$ and $P_{HVAC}$ Trajectories}

Figure~\ref{fig:Temp_PHVAC_SM1_SM2} shows the stepwise time trajectories of the zone temperature $T_{z}$ and the 3‑hour‑averaged HVAC power $P_{HVAC}$ for both SM‑1 and SM‑2 models across Spring, Winter, and Fall testing seasons at setpoints of 18\textdegree{}C and 26\textdegree{}C. Since the raw $P_{HVAC}$ signal is binary (on/off), we aggregate it over 3‑hour intervals, while $T_{z}$ is shown at every time step. In the figure, the first two rows present $T_{z}$ for SM‑1 and SM‑2, respectively, and the last two rows present the corresponding $P_{HVAC}$ profiles; within each row, subplots progress left to right through combinations of model, season, and setpoint. These results are based on parameters learned from a 21‑day summer training period at a 22\textdegree{}C setpoint and evaluated over 30‑day periods in the other three seasons (with representative 7‑day windows shown at 18\textdegree{}C and 26\textdegree{}C), omitting the summer‑22\textdegree{}C case due to uniformly high performance.

The \textbf{first two rows} present temperature trajectories for SM-1 and SM-2, while the \textbf{last two rows} show the corresponding PHVAC trajectories. \textbf{From left to right}, each subplot represents a specific combination of system model (SM-1 or SM-2), testing month (season), and setpoint (18$^\circ$C or 26$^\circ$C). The model type is indicated in each subplot, the month denotes the testing period, and the temperature value shows the setpoint at which testing was performed. The results shown here are based on models trained using 21 days of summer at setpoint of  22$^\circ$C and tested across 30 days of the remaining three seasons at setpoints of  18$^\circ$C and  26$^\circ$C, with representative trajectories shown here for a selected 7-day testing window.We have not shown it for summer and 22 setpoint as it performs good. Increasing the training dataset size beyond 21 days did not yield significant improvements in predictive performance, while greatly increasing computational cost and often leading to convergence issues in the optimization.
Among the three testing seasons, accuracy is highest in Fall and Spring and lowest in Winter as prediction is most challenging in Winter, likely due to greater differences in weather patterns and system dynamics compared to the Summer training period.For setpoint variation, prediction accuracy is similar for both 26$^\circ$C and 18$^\circ$C. Similar type of results have been generated for fall, winter and spring can be cretaed and give ismila rresults and we dont show them due to space constraint

The first two rows in Fig. \ref{fig:Temp_PHVAC_SM1_SM2} display $T_{z}$ trajectories for SM-1 and SM-2, while the last two rows show the corresponding $P_{HVAC}$ profiles. Each subplot, moving left to right, represents a combination of system model, testing season (month), and setpoint (18°C or 26°C). The results are based on models trained for 21 days in summer at a 22°C setpoint, and tested over 30 days in the remaining three seasons at 18°C and 26°C setpoints, with representative 7-day windows shown. Results for the summer 22°C case are omitted, as performance is consistently high. 

Among the three testing seasons, accuracy is highest in Fall and Spring and lowest in Winter, likely due to greater seasonal differences from the summer training data. Prediction accuracy is similar for both setpoints, and similar results were found for other seasonal and setpoint training combinations, but are omitted here due to space constraint and are instead summarized using bar plots for $T_{z}$ and $P_{HVAC}$ accuracy across all training seasons, testing periods, and setpoints as shown in Fig.\ref{fig:TEMP_PHVAC_Accuracy_2x1}. 

$T_{z}$ prediction accuracy as shown in Fig.\ref{fig:TEMP_Train_Test_Period} remains consistently high across all training and testing period combinations, with only slight differences observed between the setpoints. This shows that the model maintains strong performance for $T_{z}$ across seasons and setpoint shifts. In contrast, as visible from Fig.\ref{fig:PHVAC_Train_Test_Period}, $P_{\mathrm{HVAC}}$ accuracy shows greater variability, especially during winter testing periods where accuracy is often lower. The original (22°C) setpoint typically gives the best performance for $P_{HVAC}$, indicating that models trained at the original setpoint perform well for both $T_{z}$ and $P_{HVAC}$ trajectories but some reduction in accuracy can occur under different conditions. These results indicate that $T_{z}$ model is largely invariant to seasonal and setpoint changes, whereas $P_{HVAC}$ model is strongly influenced by variations in setpoint. Consequently, improving the accuracy of the combined $T_{z}$–$P_{HVAC}$ model will require expanding the training dataset, particularly to include a wider range of temperature setpoints.

\subsubsection{Effect of Estimation Algorithms and Reduced Order Model Types on Accuracy}

Table~\ref{tab:all_accuracy} shows that all three estimation algorithms yield high $T_{z}$ prediction accuracy, with BE and MLE producing similar results ($88.67\%$ and $88.35\%$, respectively) and LS performing slightly lower ($89.72\%$). For $P_{\mathrm{HVAC}}$, BE achieves the highest accuracy ($81.40\%$), while LS and MLE yield $80.14\%$ and $77.58\%$, respectively. Comparing reduced-order model types, both SM1 and SM2 provide comparable $T_{z}$ accuracy ($89.00\%$ and $88.83\%$), with SM1 giving slightly better $P_{HVAC}$ accuracy ($80.98\%$ vs $78.43\%$). These results suggest that the choice of estimation algorithm and model order can have a moderate effect on accuracy, particularly for $P_{HVAC}$ prediction. Thus, SM-1 paired with the BE algorithm strikes the best balance between $T_{z}$ and $P_{HVAC}$ prediction accuracy in our study. SM-1 offers a simpler model structure, which simplifies parameter estimation without sacrificing predictive performance, while BE enhances robustness by probabilistically accounting for process and sensor noise, and does so with a simpler optimization framework compared to MLE.

\begin{table}[ht]
\centering
\caption{Prediction accuracy (\%) for $T_\mathrm{z}$ and $P_{HVAC}$ by estimation method, system model, and temperature setpoint.}
\label{tab:all_accuracy}
\begin{tabular}{llcc}
\hline
Group & Name & $T_\mathrm{z}$ accuracy (\%) & $P_\mathrm{HVAC}$ accuracy (\%) \\
\hline
\multicolumn{4}{l}{\textit{Estimation method}} \\
& LS  & 89.72 & 80.14 \\
& BE  & 88.67 & 81.40 \\
& MLE & 88.35 & 77.58 \\
\hline
\multicolumn{4}{l}{\textit{Reduced-Order model}} \\
& SM1 & 89.00 & 80.98 \\
& SM2 & 88.83 & 78.43 \\
\hline
\multicolumn{4}{l}{\textit{Training season}} \\
& Fall   & 88.76 & 77.65 \\
& Spring & 89.15 & 80.15 \\
& Summer & 88.83 & 79.70 \\
& Winter & 88.92 & 81.32 \\
\hline
\multicolumn{4}{l}{\textit{Testing season}} \\
& Fall   & 88.54 & 72.67 \\
& Spring & 90.11 & 81.27 \\
& Summer & 90.03 & 89.01 \\
& Winter & 86.98 & 75.86 \\
\hline
\multicolumn{4}{l}{\textit{Temperature setpoint}} \\
& High   & 88.20 & 78.83 \\
& Low    & 89.23 & 80.66 \\
& Normal & 89.32 & 79.63 \\
\hline
\end{tabular}
\end{table}

Figure~\ref{fig:TEMP_PHVAC_Accuracy_2by2} presents bar plots of percentage accuracy for both temperature and PHVAC predictions, comparing the impact of estimation method and system model. Percentage accuracy is calculated as $100 - \mathrm{MAPE}$, providing an intuitive measure of prediction quality. While we also computed traditional error metrics such as RMSE, MAE, and MAPE, the results conveyed similar interpretation as percentage accuracy across all scenarios. For clarity and ease of comparison, we present only percentage accuracy in the accuracy index visualizations.

Interestingly, the performance comparison across estimation methods reveals that LS achieves the highest accuracy for temperature prediction, whereas MLE outperforms LS and BE for PHVAC prediction. This suggests that LS is particularly effective for direct temperature estimation, while MLE provide more robust results for modeling HVAC power consumption. The choice of system model (SM-1 vs. SM-2) leads to only minor differences in accuracy, confirming that the parameter estimation and model structure are robust across various configurations and conditions.

\subsubsection{Robustness towards Training-Testing Seasons and Setpoint Variations}

As shown in Table~\ref{tab:all_accuracy}, the proposed approach maintains robust prediction accuracy across seasons and setpoints. $T_{z}$ accuracy remains consistently high (86.98\%–90.11\%), especially when training and testing seasons align—most notably in spring and summer. $P_{HVAC}$ prediction shows greater variability, performing best during summer testing (89.01\%) and worst in fall (72.67\%), reflecting $P_{HVAC}$ usage patterns in Gainesville, FL, where higher cooling demand in spring and summer provides richer system dynamics, while fall and winter offer limited excitation. Accuracy also remains strong across setpoints, with the original 22$^\circ$C—used in training—yielding the best results (89.32\% for $T_z$ and 79.63\% for $P_{HVAC}$), and slightly lower but reliable performance at 18$^\circ$C and 26$^\circ$C. The 22$^\circ$C setpoint strikes an optimal balance, avoiding the extremes of frequent cycling or minimal $P_{HVAC}$ operation. Overall, training with data from seasons offering higher HVAC activity at a moderate setpoint offers the best combination of HVAC excitation and thermal stability, leading to more accurate models.

As summarized in Table~\ref{tab:all_accuracy}, temperature prediction accuracy remains consistently high across different training and testing seasons, with only small variations observed. Accuracy is highest when the training and testing seasons are similar, particularly in spring and summer. For $P_{\mathrm{HVAC}}$, more variability is observed, with the highest accuracy in summer testing ($89.01\%$) and the lowest in fall training ($77.65\%$) and testing ($72.67\%$). Accuracy across temperature setpoints also remains high, with the original (22$^\circ$C) setpoint giving the best results for both temperature and $P_{\mathrm{HVAC}}$, followed closely by the low (18$^\circ$C) and high (26$^\circ$C) setpoints. These results indicate that the proposed approach maintains robust prediction accuracy across a wide range of seasonal and setpoint conditions.



\section{Conclusion}\label{sec:Conclusion}

In this work, we systematically compared NLS, BE, and MLE for estimating parameters of RC-network thermal models for houses, utilizing two reduced-order modes SM-1, and SM2 to predict indoor temperature ($T_z$) and HVAC power ($P_{HVAC}$) trajectories across diverse seasons and temperature setpoints. Our results illustrated that BE with SM-1 offers a robust and balanced approach to estimate highly robust and accurate $T_z$ and $P_{HVAC}$ models. Moreover, we found that $T_z$ models are agnostic while $P_{HVAC}$ models are affected by the seasonal and setpoint variations; such that seasons for high HVAC activity with a moderate setpoint provide the best training data for estimating combined robust models. Our work serves as a practical guide for choosing estimation method, reduced-order model, and training data for developing RC-network based house thermal models which robust to seasonal and setpoint variation, as required for grid-edge applications. For future work, we will use real house data and scientific machine learning to train on larger datasets.


\section*{Acknowledgment}
This work is funded by the NSF award 2208783.


\bibliographystyle{IEEEtran}
\bibliography{BiBFile_1}


\end{document}